\documentclass[3p,times]{elsarticle}

\usepackage{ecrc}

\volume{00}

\firstpage{1}

\journalname{Physics Letters B}

\runauth{}

\jid{PLB}

\jnltitlelogo{Physics Letters B}

\CopyrightLine{2011}{Published by Elsevier Ltd.}

\usepackage{amssymb}
\usepackage{amsthm}

\usepackage[figuresright]{rotating}

\newcommand\nn{\nonumber}
\newcommand\ba{\begin{eqnarray}}
\newcommand\ea{\end{eqnarray}}
\newcommand\alb{\begin{align}}
\newcommand\ale{\end{align}}
\newcommand\be{\begin{equation}}
\newcommand\ee{\end{equation}}
\newcommand{\br}[1]{\left( #1 \right)}

\begin{document}

\begin{frontmatter}


 \title{Antiproton--proton annihilation into pion pairs within effective meson theory}


\author[label1]{Y. Wang  \corref{cor1}} 
\author[label2]{Yu. M. Bystritskiy}
\author[label3]{E.~Tomasi-Gustafsson}
\ead{etomasi@cea.fr}
\cortext[cor1]{Chinese CSC Scholar}
\address[label1]{Univ Paris-Sud, CNRS/IN2P3, Institut de Physique Nucl\'eaire, UMR 8608, 91405 Orsay, France}
\address[label2]{Joint Institute for Nuclear Research, Dubna, Russia}
\address[label3]{CEA,IRFU,SPhN, Saclay, 91191 Gif-sur-Yvette, France}

\begin{abstract}
Antiproton--proton annihilation into light mesons is revisited  in the few GeV energy domain, in view of a global description of the existing data. An effective meson model is developed, with mesonic and baryonic degrees od freedom in $s$, $t$, and $u$ channels. Regge factors are added to reproduce the proper energy behavior and the forward and backward peaked behavior. A comparison with existing data and predictions for angular distributions and energy dependence are done for charged and neutral pion pair production.
\end{abstract}

\begin{keyword}



\end{keyword}

\end{frontmatter}


\section{Introduction}
Large experimental and theoretical efforts have been going on since decades in order to understand and classify high energy processes driven by strong interaction. We revisit here hadronic reactions at incident energies above 1~GeV, focusing on two body processes. 

Antiprotons are a very peculiar probe, due to the fact that scattering and annihilation may occur in the same process, with definite kinematical characteristics. We discuss the annihilation reaction of antiproton-proton into two charged or neutral pions and  the crossed channels  of pion-proton elastic scattering. These reactions have been studied in the past, mostly at  lower energies, in connection with data from LEAR and FermiLab (for a review, see \cite{Dover:1992vj}). At low energies the annihilation into light meson pairs is dominated by few partial waves and the angular distributions show a series of oscillations. Data are usually given in terms of Legendre polynomials. This regime was studied with the aim to look for resonances in the $\bar p p$ system. A change of behavior appears around $\sqrt{s}$=2 GeV where two body  processes become mostly peripheral. They are peaked forward or backward, corresponding to small values of $t$ or $u$, respectively ($s$, $t$ and $u$ are standard kinematical Mandelstam variables of binary process). The differential cross section at large momentum transfer and the integrated cross section show a power-law behavior as a function of the energy. At larger energies, the total cross section becomes asymptotically constant and reaches a regime where $d\sigma/dt$ is function only of $t$ and is independent on $s$.  

The most exhaustive data on neutral pion (and other neutral meson) production have been published by the FermiLab E760 collaboration in the energy range ($2.2 \le \sqrt{s} \le 4.4$) GeV\cite{Armstrong:1997gv}. Charged pion production data are scarce, and do not fill with continuity a large angular or energy range \cite{Eisenhandler:1975kx,Buran:1976wc,Eide:1973tb}. According to  the foreseen performances of the experiment PANDA at FAIR, a large amount of data related to light meson pairs production from $\bar p p$ annihilation is expected in next future. The best possible knowledge of light meson production  is also requested prior to the experiment,  as pions constitute an important background for many other channels making timely the development of a reliable model. 

Few calculations exist in the literature. A baryon ($N$ and $\Delta$) $t$-channel exchange model was developed by  \cite{Moussallam:1984zj} with applicability below  1 GeV  beam momentum. A model was recently developed at larger energies, including meson exchanges in $s-$channel, which qualitatively reproduces selected sets of angular distributions \cite{VandeWiele:2010kz}. However, the authors warns against application to neighboring energies, eventually related to a specific  extrapolation of Regge trajectories in the region  $t<0$. 

We develop here a model with meson and baryon exchanges in $s$, $t$, and $u$ channels, applicable in the energy range ($ 2\le \sqrt{s} \le 5.5 $) GeV , that is the accessible  domain for the PANDA experiment at FAIR. It is known that first order Born diagrams give cross sections much larger than measured, as Feynman diagram assume point-like particles. Form factors are added in order to take into account the composite nature of the interacting particles at vertexes. Their form is, however,  somehow arbitrary, and parameters as coupling constants or cutoff are adjusted to reproduce the data.  A "Reggeization" of the trajectories is added to reproduce the very forward and very backward scattering angles. Therefore this class of models should be considered as an effective way to take into account microscopic degrees of freedom and quark exchange diagrams.

Our aim is to build a model with minimal ingredients, to calculate the basic features of neutral and charged pion production in the energy range that will be investigated by the future experiment PANDA at FAIR.  To get maximum profit  from the available data, we consider also  existing  $\pi^{\pm}p$ elastic scattering data, and apply crossing symmetry in order to compare the predictions based on the annihilation channel, at least in a limited kinematical range. The main requirement is that the model should be able to reproduce charged and neutral pion production from annihilation, and  $\pi^{\pm}p$ elastic scattering  without readjustment of the parameters.

\section{Formalism}

\subsection{Kinematics and cross section}

We consider the annihilation reaction:
\be
\bar p (p_1)+p (p_2) \to \pi^-(k_1) +\pi^+(k_2).
\label{Eq:eq1}
\ee
in the center of mass system (CMS). The  notation of four momenta is shown in the parenthesis. The following notations are used: $q_t=(-p_1+k_1)$, $q_t^2=t$, $q_u=(-p_1+k_2)$, $q_u^2=u$ and $q_s=(p_1+p_2)$, $q_s^2=s$, $s+t+u=2M_N^2+2m_\pi^2$, $M_N$($m_\pi$) is the nucleon(pion) mass. The useful scalar product between four vectors are explicitly written as:
\ba
&2p_1 k_2=2k_1p_2=M_N^2+m_\pi^2-u,\ 
&2p_1 k_1=2k_2p_2=M_N^2+m_\pi^2-t,\nn\\
&2p_1 p_2=s -2M_N^2,\ 
&2k_1k_2=s -2m_\pi^2, \nn\\
&p_1^2=p_2^2=M_N^2=E^2-|\vec p|^2, \ 
&k_1^2=k_2^2=m_\pi^2=\varepsilon^2-|\vec k|^2. \ 
\ea
The general expression for the differential cross section in the CMS of reaction (\ref{Eq:eq1}) is:
\be
\displaystyle\frac {d\sigma}{d\Omega}= 
\displaystyle\frac{1}{2^8\pi^2 } \displaystyle\frac{1}{s}
\displaystyle\frac{\beta_{\pi} }{\beta_p }|\overline{\cal M}|^{2},\ \ 
\displaystyle\frac{d\sigma}{d\cos\theta}= 2E^2\beta_p\beta_{\pi} \displaystyle\frac{d\sigma}{dt}, 
\label{Eq:tcs}
\ee
where ${\cal M}$ is the amplitude of the process, $\beta_{p} $($\beta_{\pi}) $ is the velocity and $E(\varepsilon)$ is the energy of the proton(pion) in CMS. The phase volume can be transformed as 
$d\Omega \to 2\pi \ d\!\cos\theta$ due to the azimuthal symmetry of binary reactions.
The total cross section  is :
\be
\sigma=\int \frac{|\overline{\cal M}|^2}{64 \pi^2 s} \frac {|\vec p|}{|\vec k|}d\Omega , 
\label{eq:stot}
\ee
where $|\vec p|$ and $|\vec k|$ are the initial and final momenta (moduli) in CMS.

\subsection{Crossing symmetry}
Crossing symmetry relates annihilation and scattering cross sections. Crossing symmetry states that the amplitudes of the crossed processes  are the same. This means that  the matrix element ${\cal M}(s,t)$ is the same at corresponding $s$ and $t$ values,  but the variables span different regions of the kinematical space. In order to find this correspondence, kinematical replacements between variables should be done, 
as  indicated in Table \ref{cross-kin}.
Note that the coefficients 1/2 and 1/4 in the cross section formulas are the spin factors:
$(2S_{\pi}+1)(2S_p+1)$ and 
$ (2S_{\bar p}+1)(2S_p+1)$ for the scattering and annihilation channels, respectively, where $S_\pi$, $S_p$ and $S_{\bar p}$ are the spin moduli of the corresponding initial particles. The incident momentum in the annihilation channel, corresponding to the invariant $s$ is:
$|\vec p_a|= \sqrt{{s}/{4}-M^2}$. From the equality $s_a=s_s$, the CMS momentum for $\pi^- p$ scattering, $|k_s|$, is evaluated at the same $s$ value:
\be
|\vec k_s|^2=\displaystyle\frac{1}{4s}\left [m^4-2m^2(M^2+s)+(M^2-s)^2\right ].
\label{eq:ks1}
\ee
Then the cross sections for the two crossed processes are related by: 
\be
\sigma^a=\displaystyle\frac{1}{2}\displaystyle\frac{|\vec k_s|^2}{|\vec p_a|^2} \sigma^s.
\label{eq:factor}
\ee
If the scattering cross section is measured at a value $s^s=s_1$ different from $s^a=s$, at small $t$ values  one may rescale the cross section, using the empirical dependence: $\sigma^s\simeq const\cdot s^{-2}$. 
\begin{table}
\begin{center}
\begin{tabular}{|c|c|}
\hline\hline
 Annihilation & Scattering\\
\hline
$\bar p (p_1) + p(p_2) \to \pi^-(k_1) + \pi^+(k_2)$
&
$\pi^-(-k_2)+ p(p_2) \to \pi^-(k_1)+ p (-p_1)$
\\
\hline
$s_a=(p_1+p_2)^2$ & $s_s=(-k_2+p_2)^2$\\
$t_a=(p_1-k_1)^2$ & $t_s=(-k_2-k_1)^2$ \\
$u_a=(p_1-k_2)^2$ & $u_s=(p_1-k_2)^2$~ \\
$s_a=4E^2=4(M^2+|\vec p_a|^2)$ & 
$s_s=m^2+M^2+2E_2'\epsilon_2'+2 |k_s|^2$\\
$\sigma^{(a)}= 
\displaystyle\frac{1}{4}
\displaystyle\frac
{|{\cal M}_{(a)}|^2}{64 \pi^2s} 
\displaystyle\frac{|\vec {k_a}|}{|\vec {p_a}|}$
&
$\sigma^{(s)}= \displaystyle\frac{1}{2}
\displaystyle\frac{|{\cal M}_{(s)}|^2 }
{64 \pi^2 s} 
\displaystyle\frac{|\vec k_s|}{|\vec p_s|}$\\
\hline\hline 
\end{tabular}
\caption{Correspondence between variables in the crossed scattering (s) and annihilation (a) channels.}
\label{cross-kin}
\end{center}
\end{table}

\section{Formalism}
The formulas written above are model independent, i.e., they hold for any reaction mechanism. In order to calculate 
${\cal M}$, one needs to specify a model for the reaction. In this work we consider the process (1) within the formalism of effective meson lagrangians.

\begin{figure}[htp!]
\begin{center}
\includegraphics[width=8cm]{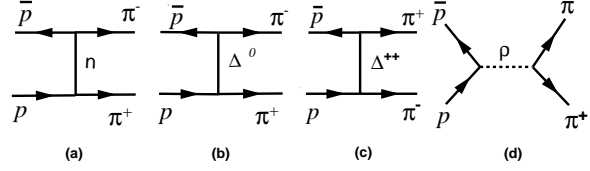}
\caption{Feynman diagrams for the reaction $\bar p+p \to \pi^- +\pi^+$ within effective meson Lagrangian approach.}
\label{Fig:DiaAll}		
\end{center}
\end{figure}

The following contributions to the cross section for reaction (\ref{Eq:eq1}) are calculated, as illustrated in Fig. (\ref{Fig:DiaAll}):
\begin{itemize}
\item  baryon exchange:
\begin{itemize}
\item  $t$-channel nucleon (neutron) exchange, Fig. \ref{Fig:DiaAll}.a,
\item  $t$-channel $\Delta^0$  exchange,  Fig. \ref{Fig:DiaAll}.b,
\item  $u$-channel $\Delta^{++}$  exchange , Fig. \ref{Fig:DiaAll}c;
\end{itemize}
\item $s$-channel  $\rho$-meson exchange,  Fig. \ref{Fig:DiaAll}d.
\end{itemize}

The total amplitude is written as a coherent sum of all the amplitudes:
\be
{\cal M}={\cal M}_n + {\cal M}_{\Delta^0} + {\cal M}_{\Delta^{++}} + {\cal M}_{\rho}.
\label{Eq:eqAll}		
\ee
In case of charged pions, the dominant contribution  in forward direction is $N$ exchange, whereas $\Delta^{++}$ mostly contribute to backward scattering.  We neglect the difference of masses between the nucleons as well as between  different charge states of the pion and of the  $\Delta$. Large angle scattering is driven by s-channel exchange of vector mesons, with the same quantum numbers as the photon. We limit  our considerations to $\rho$-meson exchange.
The expressions for the amplitudes and their interferences are detailed in the Appendix. 
Coupling constants are fixed from the known decays of the particles if it is possible, otherwise we use the values from effective potentials as \cite{PhysRevC.63.024001}. Masses and widths are taken from \cite{Agashe:2014kda}.

The effects of strong interaction in the initial state coming from the exchange by vector and (pseudo)scalar
mesons between proton-antiproton are  essential and effectively lead to the Regge form of the amplitude. The $t$ and $u$ diagrams are modified by adding a general Regge factor $R_x$ (where $x=t,u$) with  the following form:
\be
  R(x)= \left (\displaystyle\frac{s}{s_0}\right )^{2[\alpha(x)-1]},~
    \alpha(x) =  \displaystyle\frac{1}{2}+r\displaystyle\frac{\alpha_s}{\pi}\displaystyle\frac{x-M^2}{M^2};
    \label{eq:eqrt}
\ee
where $s_0\simeq 1$ GeV$^2$ can be considered a fitting parameter \cite{Kaidalov:2001db} and $r\alpha_s/\pi\simeq 0.7$ is fixed by the slope of the Regge trajectory. In the present model the values have been set at $s_0= 1.4$ GeV$^2$  and $r\alpha_s/\pi = 0.7$ for the nucleon.

This Regge form of amplitude incorporates in principle infinite number of resonances,
(i.e. $\Delta(1232)$ and others). The trajectory  for the $\Delta$ resonance is known to be different from the nucleon. 
The slope parameter is fixed in this case as $r\alpha_s/\pi = 1.4$. As for excited resonances like $N^*\br{1440}$ they belong to a daughter Regge trajectory which is power suppressed compared to the leading one. Omitting these contributions involves an estimated uncertainty of 10\%.

A form factor of the form:$F=1/(x-p^2_{N,\Delta})^2$, was introduced in the $Np\bar p$ and  $Np\bar \Delta$ vertexes, with $p_N=0.8$ GeV and $p_{\Delta}$ =5 GeV. 

The  $\rho NN$ vertex  includes the proton structure in the vector current form with two form factors (FF) $F_{1,2}^\rho$:
\be
\Gamma_\mu(q_s)= F_1^{\rho}(q_s^2) \gamma_\mu(q_s)+ \displaystyle\frac{i}{2M_N} F_2^{\rho}(q_s^2),
\sigma_{\mu\nu}q_s^{\nu},
\label{Eq:Frho}
\ee
where 
$\sigma_{\mu\nu}=  \displaystyle\frac{i}{2} [ \gamma_\mu  \gamma_\nu - \gamma_\nu  \gamma_\mu]$ is the antisymmetric tensor. Due to the isovector nature of the $\rho$, the $\rho NN$ is similar to the electromagnetic vertex $ \gamma NN$. However the two form factors 
$F_{1,2}^{\rho}(q_s^2)$ are different from the proton electromagnetic ones.
Due to the freedom of the choice, we do not attempt any rearrangement, but prefer to fix the form, the constants and the parameters of  $F_{1,2}^\rho (s)$ according to  \cite{Kuraev:2010ca,FernandezRamirez:2005iv,PhysRevC.63.024001} as:
\ba
    F_1^\rho(s) = g_{\rho NN}\frac{\Lambda^4}{\Lambda^4+(s-M_\rho^2)^2},
    \qquad
    F_2^\rho(s) = \kappa_\rho F_1^\rho(s),
 \label{eq:FFVpp}
\ea
with normalization $F_1^\rho(M_\rho^2) = g_{\rho NN}$, where the constant $g_{\rho NN}$ corresponds to the coupling
of the vector meson $\rho$ with the nucleon ($g^2_{\rho NN}/(4\pi)=0.55$) , $\kappa_\rho=3.7$ is the anomalous magnetic moment
of the proton with respect to the coupling with $\rho$, and $\Lambda=0.911$ is an empirical cut-off. 

To take into account the composite nature of the pion, in principle, a monopole type $\rho \pi\pi$ form factor may be  introduced:
$F_{\rho \pi\pi}= A_\pi/{(s-A_\pi)}^2$, where $A_\pi$ is a parameters to be adjusted on the data. In the present case  $F_{\rho \pi\pi }$ was set to one.



\begin{figure}[htp!]
\begin{center}
\includegraphics[width=10cm]{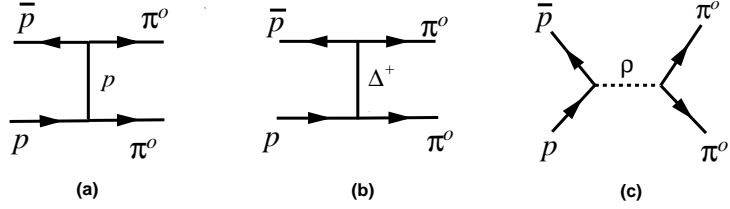}
\caption{Feynman diagrams for different exchanged particles for the reaction $\bar p+p \to \pi^0 +\pi^0$.}
\label{Fig:Diapi0}		
\end{center}
\end{figure}

The diagrams for neutral pion pair production are illustrated in Fig.  \ref{Fig:Diapi0} , where  we consider proton and $\Delta^+$ exchange in $t-$channel and $\rho$-meson exchange in $s-$ channel. The calculated amplitudes are symmetrized, to take into account the identity of the final particles.  

\section{Comparison with existing data and Discussion}
The following procedure was applied, in order to reproduce the collected data basis. The data, from Ref. \cite{Armstrong:1997gv}  on neutral pion angular distributions, were first reproduced at best, with particular attention to the $s$ dependence of the cross section and the parameters were fixed. Besides the form factors listed above, we introduced 10\% renormalization $N_ \rho$ and a mild $s$-dependent relative phase $\phi_{\rho}=\phi_0+ \phi_1 s $ of the $\rho$ diagram, with $\phi_0=1$  and $\phi_1=0.004$. 
The necessary number of parameters is very limited and we checked that the results are quite stable towards a change of the parameters in a reasonable interval.

\begin{figure}[htp!]
\begin{center}
\includegraphics[width=10cm]{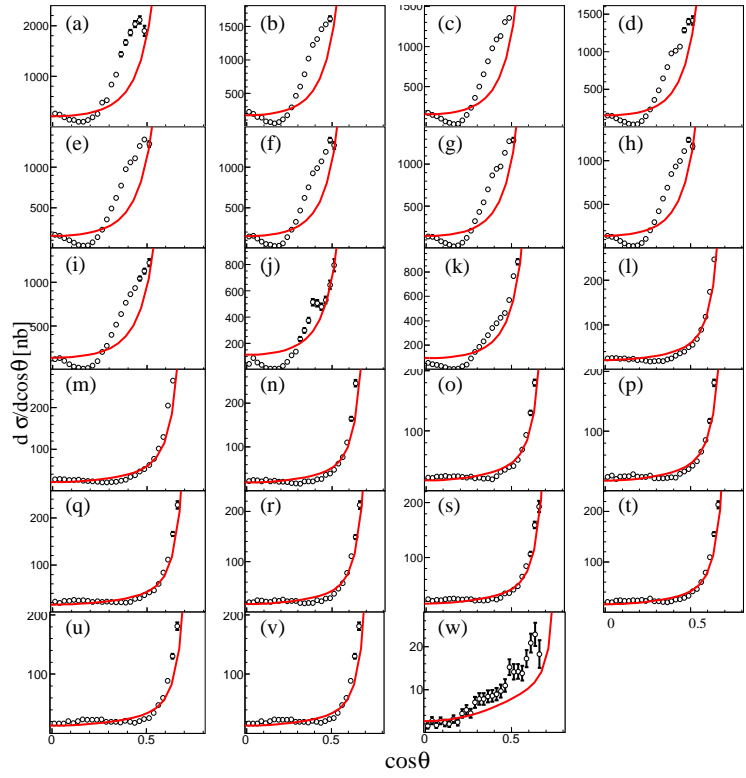}
\caption{Angular distributions for the reaction $\bar p+p \to \pi^0 +\pi^0$ for different values of   $\sqrt{s}$:
(a) $\sqrt{s}= 2.911$ GeV, 
(b) $\sqrt{s}= 2.950$ GeV, 
(c) $\sqrt{s}= 2.975$ GeV, 
(d) $\sqrt{s}= 2.979$ GeV, 
(e) $\sqrt{s}= 2.981$ GeV, 
(f) $\sqrt{s}= 2.985$ GeV, 
(g) $\sqrt{s}= 2.990$ GeV, 
(h) $\sqrt{s}= 2.994$ GeV, 
(i) $\sqrt{s}= 3.005$ GeV, 
(j) $\sqrt{s}= 3.050$ GeV, 
(k) $\sqrt{s}= 3.095$ GeV, 
(l) $\sqrt{s}= 3.524$ GeV, 
(m) $\sqrt{s}= 3.526$ GeV, 
(n) $\sqrt{s}= 3.556$ GeV, 
(o) $\sqrt{s}= 3.591$ GeV, 
(p) $\sqrt{s}= 3.595$ GeV, 
(q) $\sqrt{s}= 3.613$ GeV, 
(r) $\sqrt{s}= 3.616$ GeV, 
(s) $\sqrt{s}= 3.619$ GeV, 
(t) $\sqrt{s}= 3.621$ GeV, 
(u) $\sqrt{s}= 3.686$ GeV, 
(w) $\sqrt{s}= 4.274$ GeV. 
Data are taken from Ref. \cite{Armstrong:1997gv}, the curve is the prediction of the present model.}
\label{Fig:Angpi0}		
\end{center}
\end{figure}

The existing data on neutral pion production from Ref. \cite{Armstrong:1997gv}, together with the predictions of the model, are shown in Fig. \ref{Fig:Angpi0} for energies spanning the range ($2.911\le\sqrt{s}\le 4.274$) GeV. The angular distributions are generally well reproduced at higher energies. The agreement is less satisfactory at low energies, where, in particular, an oscillation near $\cos\theta$=0 is no accounted in the present model. This could be possibly improved by  adding other $s$-channel contributions. Moreover, the used  form of  Regge parametrization is not expected to work properly at low energies. Therefore the Regge factor was set to be unity for $s<s_0$. 

\begin{figure}
\begin{center}
\includegraphics[width=6cm]{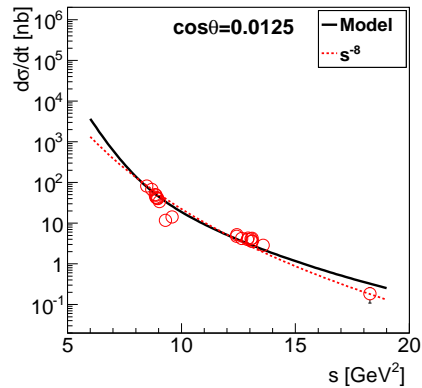}
\caption{$s$-dependence for the reaction $\bar p+p \to \pi^0 +\pi^0$ for the central region ($\cos\theta=0.0125$) and comparison of the model (black, solid line) with the $s^{-8} $ prediction from Ref. \protect\cite{Matveev:1973uz,Brodsky:1973kr} (red dashed line). Data are taken from Ref. \cite{Armstrong:1997gv}.}
\label{Fig:qcspi0}		
\end{center}
\end{figure}

The $s$ dependence of the model is shown in Fig. \ref{Fig:qcspi0},  and compared to the experimental data and to the 
$s^{-8}$ prediction from quark counting rules \cite{Matveev:1973uz,Brodsky:1973kr} for $\cos\theta=0.0125$. The model follows reasonably well the predicted behavior for large angles and large energy. A change of slope for the lower energy data is expected and was already noticed in Ref. \cite{Armstrong:1997gv}. 


Turning  to charged pion production, one more diagram  corresponding to 
$\Delta^{++}$ exchange has to be introduced in $u$ channel, to account for the asymmetric forward/backward production.   The introduction of the $\Delta^{++}$ diagram allows to reproduce the backward angles for the charged pion data. As we use the same mass and couplings for the different charged states of the $\Delta$, the same form factors parameters for $\Delta^{+,0, ++}$ are taken, not requiring any additional parameters.

\begin{figure}
\begin{center}
\includegraphics[width=10cm]{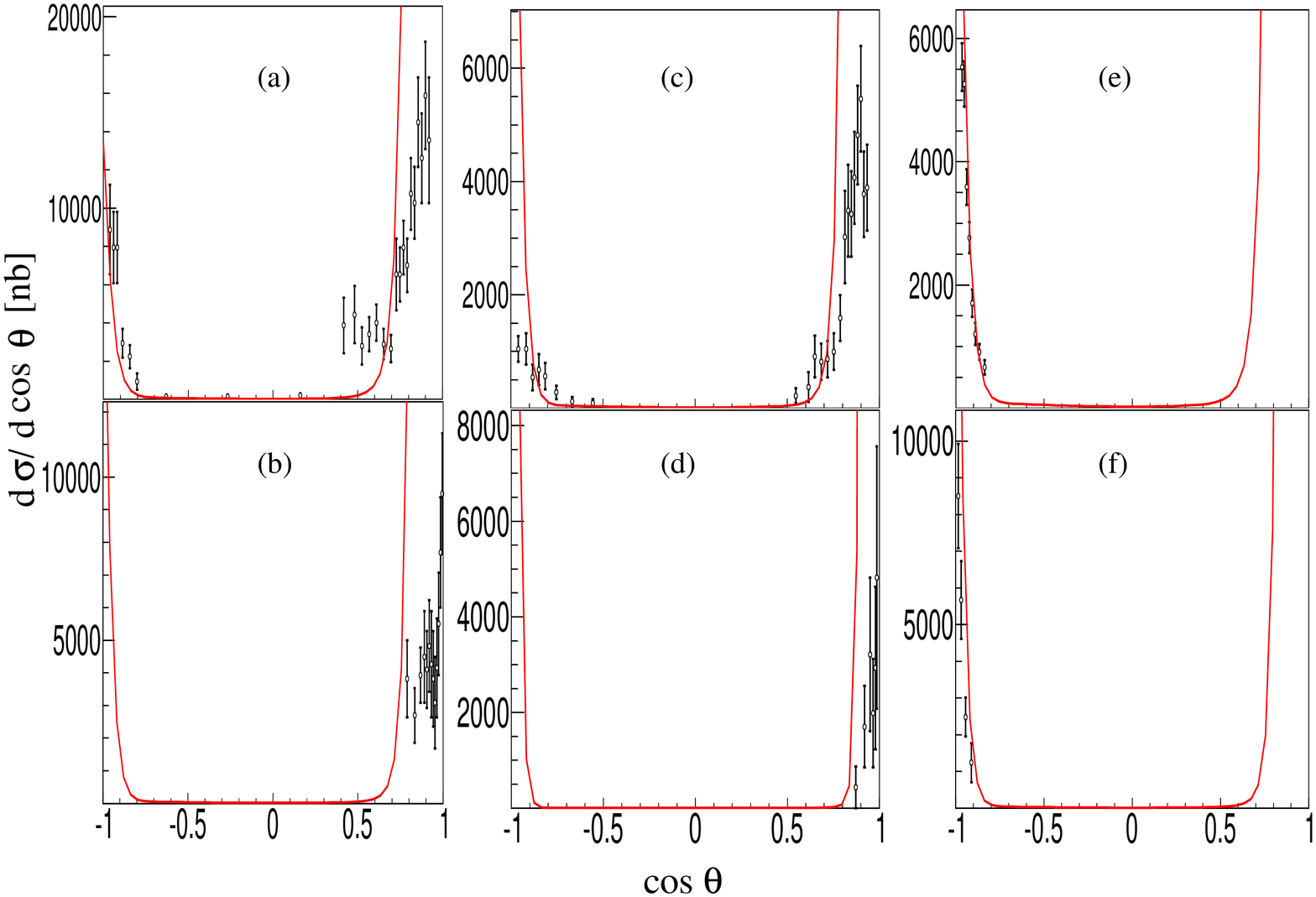}
\caption{(color online) Angular dependence for the reaction $\bar p+p \to \pi^- +\pi^+$, for different value of the total CMS energy $\sqrt{s}$ : 
(a) $\sqrt{s}=3.362$ GeV from Ref. \protect\cite{Eide:1973tb},
(b) $\sqrt{s}=3.627$ GeV from Ref. \protect\cite{Stein:1977en},
(c) $\sqrt{s}=3.680$ GeV from Ref. \protect\cite{Buran:1976wc}, 
(d) $\sqrt{s}=4.559$ GeV from Ref. \protect\cite{Berglund:1978qz}.
The corresponding data from the elastic reactions $\pi +p \to \pi +p $ are also reported: 
(e) $\sqrt{s}=3.463$ GeV from Ref. \protect\cite{PhysRev.181.1794}, 
(f) $\sqrt{s}=3.747$ GeV from Ref. \protect\cite{Baker1971385}.} 
\label{Fig:angpipm}		
\end{center}
\end{figure}

The angular dependence for the reaction $\bar p+p \to \pi^- +\pi^+$, for different value of the total CMS energy $\sqrt{s}$ are shown in Fig. \ref{Fig:angpipm} (a-d).The agreement is satisfactory, taking into account that no rearrangement of the parameters was done. They correspond to very backward angles, and are also well reproduced by the model.  

The results for the crossed channels $\pi^\pm$ elastic scattering are also reported in Fig. \ref{Fig:angpipm} (e-f), where data for the differential cross section span a small very forward or very backward angular region, bringing an additional test of the model.

\begin{figure}
\begin{center}
\includegraphics[width=6cm]{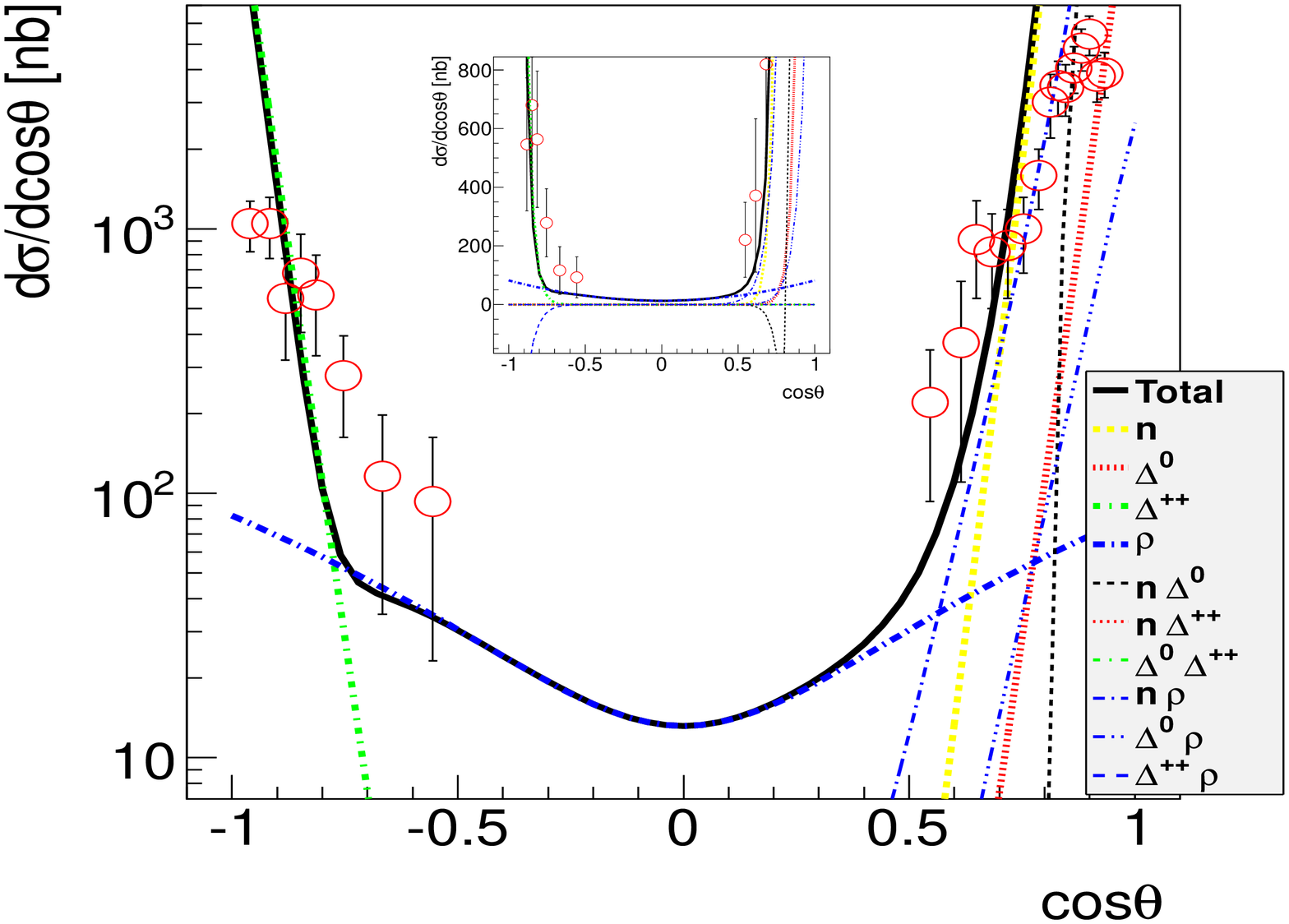}
\caption{(color online) $\cos\theta$-dependence for the reaction $\bar p+p \to \pi^- +\pi^+$ for $\sqrt{s}$= 3.680 GeV   (black, solid line) where the different components are illustrated:
$n$ exchange (yellow thick short dash line) dominates at forward angle, 
followed by $\Delta_0$( read thick dotted line)
$\Delta^{++}$ (green thick dash-dotted line) represent the largest contribution for backward angles,
$\rho$ channel  (blue thick long dash line) dominates at large angles,
The interferences are 
$n\Delta^0$ (thin black short-dash line) , 
$n\Delta^{++}$ (thin red dotted line), 
$\Delta^0 \Delta^{++}$(green thin short dash-dotted line),  
$n\rho $(blue thin long-dashed dotted line) , 
$\Delta^0 \rho $ (blue thin dash-dotted line), 
$\Delta^{++}\rho $ (blue thin long dash line).
visible at large angles. Data are taken from Ref. \protect\cite{Buran:1976wc}}. 
\label{Fig:compangpipm} 
\end{center}
\end{figure}

The angular distribution for $\sqrt{s}$= 3.680 GeV is shown in Fig. \ref{Fig:compangpipm}   The total result (black, solid line) gives a good description of the data (red open circles) from Ref. \cite{Buran:1976wc} for charged pion production. All components and their interferences are illustrated. The main contribution at central angles is given by $\rho$ s-channel exchange, whereas $n$ exchange in $t$ channel dominates forward angles followed by $\Delta^0$ exchange. $\Delta^{++}$ represent the largest contribution for backward angles. 
The interferences are also shown. Their contribution affects the shape of the angular distribution, some of them being negative in part of the angular region.

%

\section{Conclusions}
An model, built on effective meson Lagrangian, has been build in order to reproduce the existing data for two pion production in proton-antiproton annihilation at moderate and large energies. Form factors and Regge factors are implemented and parameters adjusted to the existing data for neutral and charged pion pair production. Coupling constants are fixed from the known properties of the corresponding decay channels. The agreement with a large set of data is satisfactory for the angular dependence as well as the energy dependence of the cross section. At large angles the model follows naturally the expected behavior from quark counting rules. 

A comparison with data from elastic $\pi^{\pm} p \to \pi^{\pm} p$, using crossing symmetry prescriptions  shows a good agreement also at very forward and backward angles, within the uncertainty. Discussion about validity of crossing symmetry can be found in Refs. \cite{Eide:1973tb,PhysRev.181.1794,Stein:1977en}. The present results verify  that crossing symmetry works at least at backward angles, where one diagram is dominant.

This model can be extended to other binary channels, with appropriate  changes of constants. The implementation to MonteCarlo simulations for predictions and optimization to coming experiments is also foreseen. 


\section{Acknowledgments}
Thanks are due to D. Marchand and A. Dorokhov, for useful discussions and interest in this work. One of us (Yu.B) acknowledges kind hospitality at IPN Orsay, in frame of JINR-IN2P3 agreement.

\section{Appendix}
The relevant formulas for the amplitudes and their interferences are given below.
\subsection{Neutron exchange}
The amplitude for nucleon exchange is written as: 
\be
{\cal M}_N = \displaystyle\frac {g^2_{\pi NN}}{q_t^2-M_N^2}\bar v (p_1)  (-\hat q_t+M_N) u(p_2).
\label{Eq:ampl_a}
\ee
 where $u(p_2)(\bar v( p_1))$, are the four-component spinors of the proton(antiproton), which obey the Dirac equation. 
The matrix element squared for the diagram corresponding to neutron exchange, Fig. \ref{Fig:DiaAll}a is written as:
\ba
|\overline {{\cal M}_N}| ^2 &=& \displaystyle\frac{ g^4_{ \pi NN}} {(q_t^2-M_N)} Tr \left  [ (\hat p_1-M_N)  (-\hat q_t+M_N) (\hat p_2+M_N) (-\hat q_t+M_N) \right ]\nn\\
&= &-2\displaystyle\frac{ g^4_{ \pi NN}} {(t-M_N) ^2} \left [ m^4 +(M_N^2-t)(M_N^2-s -t +2m_\pi) \right ],
\label{Eq:tr_a}
\ea
with $q_t=k_1-p_1=p_2-k_2$, $q_t^2=t$.

\subsection{$t$-Exchange of $\Delta^0$}

The specific ingredients for $\Delta$ exchange, Fig. \ref{Fig:DiaAll}b, are related to the spin 3/2 nature of the 
$\Delta$-resonance . For the $\Delta$-spin-vector, $U_{\Delta}$, we take the expression from \cite{AkhiezerRekalobook::1977}:
where the density matrix is 
\be
P_{\mu\nu}=U_\mu^\Delta(p_{\Delta})\bar U_\nu^{*\Delta}(p_{\Delta})=-g_{\mu\nu}+
\frac{1}{3} \gamma_{\mu}\gamma_{\nu}+
\frac{\gamma_{\mu} P_{\nu}-\gamma_{\nu}P_{\mu}} {3M_{\Delta}}+
\frac{2}{3}\frac{P_{\mu}P_{\nu}}{M_{\Delta}^2}.
\label{Eq:UU*}
\ee
and: 
\be
a) \displaystyle\frac{i}{(2\pi)^4}\displaystyle\frac{\hat q_t+M_{\Delta}}{q_t^2-M_{\Delta}^2} P_{\mu\nu}, ~~
b) -i(2\pi)^4 g_{\Delta\pi N}k_1^\mu 
\label{Eq:ampl_b}
\ee
are  the expressions for a) the $\Delta$ propagator and b) the vertex $\Delta\to \pi N$. 
$M_\Delta=1.232 \pm 2 $ MeV is the weighted mass of the  $\Delta$ resonance, (i.e., the mass averaged over $\Delta$-multiplet),  and  $g_{\Delta\pi N}$ is the coupling constant for the vertex $\Delta\to \pi N$.

The matrix element for the diagram Fig. \ref{Fig:DiaAll}.b is:
\be
{\cal M}_{\Delta^0}=-\displaystyle\frac{ g^2_{\Delta\pi N}} {t-M_{\Delta}^2} \bar v (p_1)(\hat q_t+M_{\Delta}) P_{\mu \nu}(q_t) u(p_2)k_1^\mu k_2^\nu.
\label{Eq:MD0}
\ee
Squaring the amplitude one has
\be
|\overline{{\cal M}_{\Delta^0}}| ^2 =\displaystyle\frac{ g^4_{\Delta\pi N}} {(t-M_{\Delta}^2) ^2} k_1^\mu k_2^\nu k_1^\alpha k_2^\beta
Tr \left  [ (\hat p_1-M_N)  (\hat q_t+M_{\Delta}) P_{\mu \nu}(q_t)
 (\hat p_2+M_N)\tilde P_{\alpha \beta}(q_t)(\hat q_t+M_{\Delta}) \right ] .
\label{Eq:tr_d0}
\ee
In order to find the value of $g_{\Delta N \pi}$ coupling constant we consider the decay width of $\Delta$ into nucleon and pion: 
\be
\Gamma_\Delta=  \displaystyle\frac{3}{2} \displaystyle\frac{|\vec p|}{32 \pi M_{\Delta}^2} 
|{\cal M}(\Delta\to  N \pi)|^2, 
\label{Eq:gamma2f}
\ee
and using the experimental values of the decay width $\Gamma_\Delta= 117\pm 3$  MeV \cite{Agashe:2014kda}, one can estimate  $g_{\Delta N\pi}= 15.7\pm 0.4 $ GeV$^{-1}$.

\subsection{$u$-exchange of $\Delta^{++}$ }
The diagram in Fig. \ref{Fig:DiaAll}c corresponds to $\pi^-$ emitted at backward angle involves the exchange of $\Delta^{++}$ and can be obtained from $t$-exchange \ref{Fig:DiaAll}b with the replacements: $t \leftrightarrow u$ and  $k_1\leftrightarrow k_2$. 
\subsection{ Interferences with $\Delta $ }
\subsubsection{ The $\Delta^0-N $ interference}
\be
2Re[M_N^* M_{ \Delta}^0]= 2Re  
\Bigg \{
\displaystyle\frac{ g^2_{\pi NN} g^2_{\Delta \pi N} } 
{ (t-M_N^2)(t-M_{ \Delta}^2)} Tr \left  [ (\hat p_1 + M_N)   (-\hat q_t+M_N)
 (\hat p_2 + M_N)     
 \tilde P_{\mu \nu}(q_t)(\hat q_u+M_{\Delta})   \right ]
k_1^\mu k_2^\nu  
\Bigg \},
 \label{Eq:intNDelta}
 \ee
 with $q_u=k_2-p_1=p_2-k_1$, $q_u^2=u$.
\subsubsection{ The $\Delta^{++}-N $ interference }
\be
2Re[M_N^* M_{ \Delta}^{++}]= 2Re  \Bigg \{
\displaystyle\frac{ g^2_{\pi NN} g^2_{\Delta \pi N} } 
{ (u-M_{ \Delta}^2)(t-M_N^2)} 
Tr \left  [ (\hat p_1 + M_N)   (-\hat q_t+M_N)
 (\hat p_2 + M_N)    \tilde P_{\mu \nu}(q_u)(\hat q_u+M_{\Delta})   
 \right ] k_1^\nu k_2^\mu  
\Bigg \}.
 \label{Eq:intNDelta++}
  \ee
\subsubsection{ The $\Delta^{++}-\Delta^0 $ interference }
\be
2Re[M_{ \Delta}^{*0} M_{ \Delta}^{++}]= 
2Re \Bigg \{
\displaystyle\frac{ g^4_{\Delta \pi N} } 
{ (t-M_{ \Delta}^2)(u-M_{ \Delta}^2)} 
\label{Eq:intDelta0Delta++}
Tr \left  [ (\hat p_1 - M_N)   (\hat q_t+M_N)  
 P_{\mu \nu}(q_t)
  (\hat p_2 + M_N)  
  \tilde P_{\alpha \beta}(q_u)  (\hat q_u+M_{\Delta})    \right ]
k_1^\mu k_2^\nu k_2^\alpha k_1^\beta  
\Bigg \}.
  \ee

\subsection{$s$-exchange of $\rho$ meson}
The largest contribution to meson exchange in s-channel, Fig \ref{Fig:DiaAll}d,  is given by the $\rho$-meson, with  $\sim 100\%$ branching ratio into two pions.
 For the a) $\rho$- propagator and b) the $\rho \pi\pi $ vertex we take  
 \be
 a) -\displaystyle\frac{i}{(2\pi)^4} \left [
 \displaystyle\frac{g_{\mu\nu}-(q_s^{\mu}q_s^{\nu})/{m_{\rho}^2}}
 {q_s^{2}-m_{\rho}^{2}+i \sqrt{q_s^2}\Gamma_\rho(q_s^2) }\right ], ~~  b) -ig_{\rho\pi\pi}(k_1-k_2)^\nu (2\pi)^4,~q_s=p_1+p_2=k_1+k_2,
 \ee
 where $g_{\mu\nu}$ is the symmetric tensor, and  $q_s^2=s$.
The matrix element is written as: 
\be
{\cal M}_{\rho} =   \displaystyle\frac { g_{\rho p p } g_{\rho \pi \pi}}{[s-m_{\rho}^{2}+i \sqrt{s}\Gamma_\rho(s) ]}
 [ \bar v (p_1)   \Gamma^\mu(q)  u(p_2)](k_1 -k_2)^\nu 
  \left\{ 
 g_{\mu\nu}-\displaystyle\frac{q_{\mu}q_{\nu}} {m_{\rho}^2}
  \right \} ,
\label{Eq:ampl_d}
\ee
Squaring the amplitude one gets:
\ba
|\overline{ {\cal M}_{\rho}}| ^2 &=&
\displaystyle\frac{ g^2_{\rho NN} g^2_{\rho \pi \pi}} 
{[ s-m_{\rho}^2+i \sqrt{s}\Gamma_\rho(s) ]^2} 
(k_1-k_2)^\nu (k_1-k_2)^\beta  
\left ( g_{\mu\nu}-\displaystyle\frac{(q_s)_\mu(q_s)_\nu}{m_{\rho}^2} \right )
\left ( g_{\alpha\beta}-\displaystyle\frac{(q_s)_\alpha (q_s)_\beta} {m_{\rho}^2} \right )
\nn\\
&& 
Tr \left  [ (\hat p_1 - M_N)  
\Gamma^\mu(q_s)  (\hat p_2 + M_N)  
\Gamma^\alpha(q_s)  \right ]. 
\label{Eq:tr_rho}
\ea
The coupling constant $g_{\rho \to \pi \pi}$ is found from the the experimental value of the  total width  $\Gamma$ for the decay $\rho \to \pi \pi$: $\Gamma(\rho)$ = 149.1 $ \pm$ 0.8 MeV \cite{Agashe:2014kda}. The branching ratio into two pions is  $\approx 100 \%$.  The total width has the form: 
\be
 \Gamma =\displaystyle\frac{4}{3}
\frac{g_{\rho\pi\pi}^2}{16 \pi m_{\rho}^2}  (m_{\rho}^2 - 4 m_\pi^2)^{3/2},
\label{Eq:phaseR}   
\ee
where we added a factor 4/3 to take into account that there are three possible initial states of the $\rho$ meson and  four possible charged decays. Inverting Eq. (\ref{Eq:phaseR}), using the experimental value for the decay width one can get the following value of the coupling constant:  
$g_{\rho\pi\pi}=5.175\pm 0.017 $.
\subsection{Interferences with $\rho$}
\subsubsection{ The $N-\rho $ interference }
\ba
2Re[M_N^*M_{\rho}]&= &
2Re \left \{
\displaystyle\frac{ g_{\pi NN } g_{\rho \pi\pi} g^2_{\rho NN}} 
{[ s-m_{\rho}^2+i \sqrt{s}\Gamma_\rho(s) ] (t-M_N^2)} 
Tr \left  [ (\hat p_1 - M_N)  
\Gamma^\mu(q_s)  (\hat p_2 + M_N) 
\tilde P_{\alpha \beta}(q_t)  (-\hat q_t+M_N)   
\right ]
\right  .\nn\\
&&  \left .
k_1^\alpha k_2^\beta
(k_1 -k_2) ^\nu  
\left (g_{\mu\nu}-\displaystyle\frac{q_{\mu}q_{\nu}}{m_{\rho}^2} \right )
\right \}.
\label{Eq:intNrho}
\ea
\subsubsection{ The $\Delta^0-\rho $ interference }
\ba
2Re[M_{\Delta^0}m_{\rho}^*]&= &
2Re \left \{
\displaystyle\frac{ g_{\rho NN } g_{\rho \pi \pi} g^2_{\Delta\rho N}} 
{[ s-m_{\rho}^2+i \sqrt{s}\Gamma_\rho(s) ] (t-M_{\Delta}^2)} 
Tr \left  [ (\hat p_1 - M_N)  
\Gamma^\mu(q_s)  (\hat p_2 + M_N) 
\tilde P_{\alpha \beta}(q_t)  (-\hat q_t+M_{\Delta})   
\right ]
\right  .\nn\\
&&  \left .
k_1^\alpha k_2^\beta
(k_1 -k_2) ^\nu  
\left (g_{\mu\nu}-\displaystyle\frac{q_{\mu}q_{\nu}}{m_{\rho}^2} \right )
\right \}
\label{Eq:intD0rho}.
\ea
\subsubsection{ The $\Delta^{++}-\rho $ interference }
\ba
2Re[M_{\Delta^{++}}m_{\rho}^*]&=&
2Re \left \{
\displaystyle\frac{ g_{\rho NN } g_{\rho \pi \pi} g^2_{\Delta\rho N}} 
{[ s-m_{\rho}^2+i \sqrt{s}\Gamma_\rho(s) ] (u-M_{\Delta}^2)} 
Tr \left  [ (\hat p_1 - M_N)  
\Gamma^\mu(q_s)  (\hat p_2 + M_N) 
\tilde P_{\alpha \beta}(q_u)  (-\hat q_u+M_{\Delta})   
\right ]
\right  .
\nn\\
&&  \left .
k_1^\alpha k_2^\beta
(k_1 -k_2) ^\nu  
\left (g_{\mu\nu}-\displaystyle\frac{q_{\mu}q_{\nu}}{m_{\rho}^2} \right )
\right \}.
\label{Eq:intD++rho}
\ea

\bibliography{Biblio}
\end{document}